\documentclass[twocolumn,pra,showpacs,superscriptaddress]{revtex4}
\usepackage{amssymb}
\usepackage{amsmath}
\usepackage{graphicx}
\usepackage{subfigure}
\usepackage{natbib}
\usepackage{epsfig}
\usepackage{amsfonts}
\usepackage{mathrsfs}
\usepackage{ulem}
\usepackage{color}
\usepackage[toc,page,title,titletoc,header]{appendix}

\begin{document}

\title{Mapping trapped atomic gas with spin-orbit coupling to quantum Rabi-like model}

\author{Haiping Hu}
\affiliation{Beijing National Laboratory for Condensed Matter
Physics, Institute of Physics, Chinese Academy of Sciences,
Beijing 100190, China}
\author{Shu Chen}
\email{schen@aphy.iphy.ac.cn} \affiliation{Beijing National
Laboratory for Condensed Matter Physics, Institute of Physics,
Chinese Academy of Sciences, Beijing 100190, China}
\begin{abstract}
We construct a connection of the ultracold atomic system in a
harmonic trap with Raman-induced spin-orbit coupling to the
quantum Rabi-like model. By mapping the trapped atomic system to a
Rabi-like model,
we can get the exact solution of the Rabi-like model following the
methods to solve the quantum Rabi model. The existence of such a
mapping implies that we can study the basic model in quantum
optics by using trapped atomic gases with spin-orbit coupling.
\end{abstract}



\maketitle

\section{Introduction}
Recently, many attentions have been paid to the cold atomic system
with synthetic gauge field and spin-orbit (SO) coupling, which
have been successfully realized in ultracold Bose gases
\cite{NIST,NIST_elec,SOC,collective_SOC,NIST_partial,JingPRA,ZhangPRA}
and Fermi gases \cite{SOC_Fermi,SOC_MIT}. The experimental
progress stimulated the intensive theoretical studies, including
exploring schemes to create general gauge fields
\cite{gauge_review,You} and studying various interesting phases in
these novel atomic systems with SO coupling
\cite{Stripe,Ho,Victor,Wu,Hu,Santos,Cui}. While many theoretical
works focused on the uniform systems with isotropic Rashba-type SO
coupling, trapped systems with an external harmonic trap have also
been addressed \cite{Wu,Hu,Santos,Ghosh,Larson}. It was shown that
the presence of a confining potential may qualitatively change the
physical properties of atomic gases with SO coupling.

In this work, we shall consider the anisotropic one-dimensional
(1D) SO coupling realized in current experiments and scrutinize
the problem of cold atomic system with SO coupling in a harmonic
trap. We find that even the single particle problem of the trapped
atomic system is highly nontrivial and not easy to be solved
analytically when both the SO coupling term and Raman-coupling
term exist. As we shall show in the context,  the trapped atomic
system with Raman-induced spin-orbit coupling is equivalent to a
quantum Rabi-like model. As a paradigm for modeling the simplest
light-atom interacting quantum system \cite{Rabi,JC}, the quantum
Rabi model \cite{Rabi} was surprisingly not able to be
analytically solved and only very recently has an analytical
solution been found by Braak \cite{Braak}. The theoretical
progress has renewed the interest in the study of the quantum Rabi
model \cite{chenqh,Ziegler,Gardas,Solano,dynamics}. Particularly,
Braak's solution can be also re-derived \cite{chenqh} within the
extended coherent stats \cite{chenqh-EPL,FengM} in a more
straightforward way.

Within the same scheme of solving the quantum Rabi model
\cite{Braak,chenqh}, we can solve the Rabi-like model and get an
analytical solution in terms of Braak's transcendental functions.
By using the analytical solution, we can calculate the energy
spectrum and dynamics of the Rabi-like model exactly. As the Rabi
model has been widely applied to different fields of physics,
including quantum optics, the cavity and circuit quantum
electrodynamics \cite{CQED,CQED2} and semiconductor systems
\cite{Semiconducor}, et al., we expect that the mapping of the
trapped atomic system with spin-orbit coupling to quantum
Rabi-like model paves the way to study physical phenomena related
to quantum Rabi model by using cold atomic systems.


\section{The trapped atomic gas with SO coupling}
 The Hamiltonian
of a single atom with SO coupling takes form
\begin{equation}
\hat{H}_{3d}=\mathbf{\hat{p}}^{2}/(2m)+
\hat{V}_{SO}+\hat{V}(\mathbf{\hat{r}}),
\end{equation}
with $\hat{V}(\mathbf{\hat{r}})=\frac{1}{2}m (\omega_x^2 x^2 +
\omega_y^2 y^2 +\omega_z^2 z^2 )$ the trap potential, $m$ the
atomic mass, and $\mathbf{\hat{p}}$ the atomic momentum operator.
The term of $\hat{V}_{SO}$ describes the SO coupling. For effective spin-$1/2$ systems considered in Refs.~%
\cite{SOC,JingPRA}, one has
\begin{equation}
\hat{V}_{SO}=2k_{r}\hat{p}_{x}\hat{\sigma}_{z} + \frac{\Omega}{2}
\hat{\sigma}_{x} + \frac{\delta}{2} \hat{\sigma}_{z}, \label{SO}
\end{equation}
where $ k_{r}$ is the recoil momentum, $\Omega $ is the
Raman-coupling strength, $\delta $ is the two-photon detuning and
$\hat{\sigma}$ represents the Pauli operators. For the SO coupling
described by Eq.(\ref{SO}), the Hamiltonian can be separated into
$\hat{H}_{3d} = \hat{H}(x) + \hat{H}_{2d}(y,z)$ with the effective
1D Hamiltonian given by
\begin{equation}
\hat{H} =\frac{\hat{p}_x^{2}}{2m} + \frac{1}{2} m \omega^2 x^2 +
\hat{V}_{SO}, \label{H1d}
\end{equation}
where we have used $\omega=\omega_x$ for brevity. The model of
(\ref{H1d}) has a deceptively simple form. However, as we shall
show in the next calculation, the model is not easy to be solved
except the special cases with either $\omega=0$ or $\Omega=0$.

\section{Mapping to the Rabi-like model }
To make a connection of
the SOC model in the harmonic trap to the well known quantum Rabi
model, we use the representation of ladder operators, i.e., $a$
and its adjoint $a^{\dagger}$, to rewrite the Eq.(\ref{H1d}) as
\begin{equation}
\hat{H} = \hbar \omega (a^{\dagger} a + \frac{1}{2}) +
\delta \hat{\sigma}%
_{z}/2+\Omega \hat{\sigma}_{x}/2+ ig
(a^{\dagger}-a)\hat{\sigma}_{z}, \label{Rabi-like}
\end{equation}
where $g=k_{r}\sqrt{2 m \hbar \omega}$ and $\hat{p}_x= i \sqrt{{m
\hbar \omega}/{2}}(a^{\dagger}-a)$ is used. To simplify the above
equation and compare with Rabi model, we make a shift $H
\rightarrow H-\hbar \omega/2$ and take $\hbar=1$. Now the
Hamiltonian can be rewritten as
\begin{equation}
\hat{H} = \omega a^{\dagger} a  + \frac{\delta}{2} \hat{\sigma}
_{z} +\frac{\Omega}{2} \hat{\sigma}_{x} + ig
(a^{\dagger}-a)\hat{\sigma}_{z}. \label{Rabi-like}
\end{equation}
We note that a generalized quantum Rabi model can be represented
as
\begin{equation}
\hat{H}_{R} = \omega a^{\dagger} a  + \frac{\delta}{2}
\hat{\sigma} _{z} +\frac{\Omega}{2} \hat{\sigma}_{x} + g
(a^{\dagger} + a)\hat{\sigma}_{z}, \label{Rabi}
\end{equation}
with $\delta=0$ corresponding to the quantum Rabi model.
Comparing Eq.(\ref{Rabi-like}) with the generalized Rabi model,
we find that the only difference is the spin-boson coupling term.
While the coupling term in the generalized Rabi model describes a
spin-space coupling,  the coupling term in the Rabi-like model of
the trapped atomic gas describes a spin-momentum coupling.

The resemblance of these two models suggests us that the Rabi-like
model (\ref{Rabi-like}) can be solved by applying similar methods
to solve the quantum Rabi model \cite{chenqh,Braak}. Here we shall
use the method of Chen et. al \cite{chenqh} to derive our solution
of Rabi-like model. For brevity, we give results and key steps of
derivation in the main text, but leave some details of derivation
in the appendix. To diagonalize the Rabi-like model, it is
instructive to span the Hamiltonian (\ref{Rabi-like}) in the spin
space with the following matrix form
\begin{equation}
\hat{H}=\left(
\begin{array}{ll}
a^{\dagger }a+i g\left( a^{\dagger } - a\right) + \frac \delta 2 &
~~~~~~~~\frac \Omega 2 \\
~~~~~~~~\frac \Omega 2 & a^{\dagger }a- i g\left( a^{\dagger }-
a\right) - \frac \delta 2
\end{array}
\right) .
\end{equation}
Here, in order to keep consistent with Ref.\cite{chenqh}, we take
the matrix form in units of $\omega=1$. By introducing the
Bogoliubov transformations $A=a+ig$ and $B=a-ig$, the diagonal
matrix element $H_{11}$ and $H_{22}$ can be diagonalized as
$A^{\dagger }A$ and $ B^{\dagger }B$, respectively.

In terms of operator $A$, we can rewrite the Hamiltonian as
\begin{equation} \hat{H}=\left(
\begin{array}{ll}
A^{\dagger }A-\alpha  & ~~~~~~~~\frac \Omega 2  \\
~\frac \Omega 2  & A^{\dagger }A-2ig\left( A^{\dagger }- A\right)
+\beta
\end{array}
\right) ,
\end{equation}
where
\[
\alpha =g^2-\frac \delta 2,\;\beta =3g^2-\frac \delta 2.
\]
To diagonalize the Hamiltonian, we take the wavefunction as
\begin{equation}
\left| { \psi }\right\rangle =\left( \sum_{n=0}^\infty
\sqrt{n!}J_n^- \left| n\right\rangle _A, \sum_{n=0}^\infty
\sqrt{n!}K_n^-\left| n\right\rangle _A \right)^T ,  \label{wave1}
\end{equation}
where $\left| n\right\rangle _A = \frac{\left( A^{\dagger }\right)
^n}{\sqrt{n!}} \left| 0\right\rangle _A$ is the extended coherent
state with $\left| 0\right\rangle _A = e^{-\frac 12g^2-ig
a^{\dagger }}\left| 0\right\rangle _a$. The expansion coefficients
$J_n^-$ and $K_n^-$ can be determined self-consistently by solving
the eigenvalue equation
 $H \left| { \psi }\right\rangle = E \left| { \psi }\right\rangle $,
which gives some restricted relations to coefficients $J_n^-$ and
$K_n^-$ \cite{chenqh} (see appendix).

Similarly, in terms of operator $B$, the Hamiltonian is written as
\begin{equation}
\hat{H}=\left(
\begin{array}{ll}
B^{\dagger }B+2ig\left( B^{\dagger } - B\right) +\beta ^{\prime }
& ~~\frac
\Omega 2 \\
~~~~~~~~~~~~\frac \Omega 2 & B^{\dagger }B-\alpha ^{\prime }
\end{array}
\right) ,
\end{equation}
where
\[
\alpha ^{\prime }=g^2+\frac \delta 2,\;\;\beta ^{\prime
}=3g^2+\frac \delta 2.
\]
The wavefunction can be also represented in terms of $B$ as
\begin{equation}
\left| {\psi'}\right\rangle =\left( \sum_{n=0}^\infty \left(
-1\right) ^n\sqrt{n!} K_n^+ \left| n\right\rangle _B,
\sum_{n=0}^\infty \left( -1\right) ^n\sqrt{n!} J_n^{+} \left|
n\right\rangle _B \right)^T . \label{wave2}
\end{equation}
where $\left| n\right\rangle _B = \frac{\left( B^{\dagger }\right)
^n}{\sqrt{n!}} \left| 0\right\rangle _B$ is the extended coherent
state with $\left| 0\right\rangle _B = e^{-\frac 12g^2+ig
a^{\dagger }}\left| 0\right\rangle _a$, and $K_n^+$ and $J_n^+$
are the coefficients to be determined by solving the eigenvalue
equation
 $H \left| { \psi' }\right\rangle = E \left| { \psi' }\right\rangle
 $.

Since both wavefunctions  $\left| {\psi}\right\rangle$ and $\left|
{\psi'}\right\rangle$ are eigenfunctions for the same eigenvalue
$E$, they should be different by a complex constant if they are
not degenerate. This requirement leads to a self-consistent
condition between the two sets of coefficients, i.e.,
\[
\sum_{n=0}^\infty J_n^- (-ig)^n\sum_{n=0}^\infty J_n^{+
}(-ig)^n=\sum_{n=0}^\infty K_n^- (-ig)^n \sum_{n=0}^\infty
K_n^{+}(-ig)^n,
\]
The above equation can be rewritten as a compact form $G_\delta
(x)=0$ with the transcendental function defined as
\begin{equation}
G_\delta (x)=(\frac \Omega 2)^2\bar{R}^{+}(x)\bar{R}
^{-}(x)-R^{+}(x)R^{-}(x),  \label{bias}
\end{equation}
where $x=E+g^2$
\begin{eqnarray*}
R^{\pm }(x) &=&\sum_{n=0}^\infty K_n^{\pm }(x)\;(-i g)^n, \\
\bar{R}^{\pm }(x) &=&\sum_{n=0}^\infty \frac{K_n^{\pm }(x)}{x-n
\pm \frac \delta 2}\;(-i g)^n.
\end{eqnarray*}
The $ K_n^{\pm }(x)$ are  defined recursively,
\begin{equation}
n  K_n^{\pm }= f_{n-1}^{\pm }(x) K_{n-1}^{\pm } + K_{n-2}^{\pm },
\label{Kn}
\end{equation}
with the initial condition $ K_0^{\pm }=1$,  $ K_1^{\pm }=
f_{0}^{\pm }(x)$, and
\begin{equation}
f_{n}^{\pm }(x) =2ig-\frac 1{2ig}\left[  n -x \pm \delta/2 -
\frac{(\Omega/2) ^2}{4\left( x- n \mp \delta/2 \right) }\right] ,
\end{equation}
The eigenenergies can be determined by the zeros of the
transcendental function $G_\delta (x)$ \cite{Braak}.  For $\delta
=0$, $G_\delta (x)=0$ can be reduced to $G_0^{\pm }(x)=0$ with
\begin{equation}
G_0^{\pm }(x)=\sum_{n=0}^\infty K_n(x)\left( 1\pm \frac{\Omega /2}{x-n}%
\;\right) (-ig)^n . \label{unbias}
\end{equation}
\begin{figure}[tbp]
\includegraphics[width=3.5in]{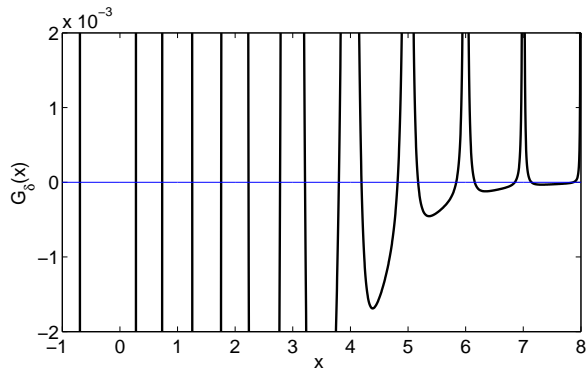}\newline
\caption{(Color online) The energy of the Rabi-like model can be
determined from the points of intersection of G-function. Here we
set $\Omega=1.4$, $g=0.1$, $\delta=0$ and made a truncation of
first 199 coherent state orbits.}
\end{figure}

\section{Results and discussions}
The spectrum of the Rabi-like model can be determined from the
zeros of transcendental function (\ref{bias}) or (\ref{unbias}).
As illustrated in Fig.1, we can read the eigenenergy from the
intersection points of G-function with horizontal axis ($x_{in}$)
via $E=x_{in}-g^2$. Using the above G-function, it is quite easy
to get the relation of energy spectrum as function of coupling
constant. To give a concrete example, we plot the energy spectrum
for the system with $\Omega=1.4$ in Fig.2. As shown
in the figure, the ground state energy decreases with increasing
coupling constant $g$. When $\delta/2$ is not a multiple of $\omega/2$,
 no level crossings appear. This is consistent with Braak's  results \cite{Braak}.

\begin{figure}[tbp]
\includegraphics[width=3.6in]{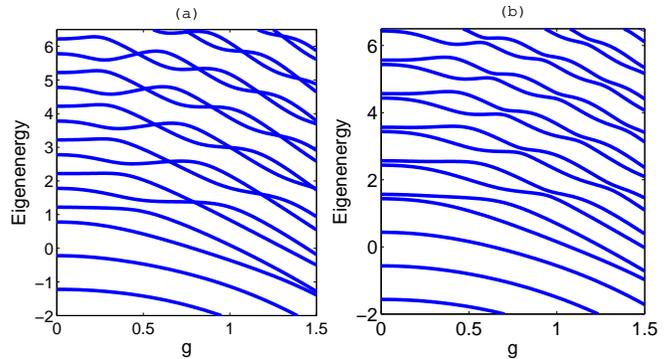}
\caption{(Color online) The energy of the Rabi-like model versus
coupling strength $g$. Here we set $\Omega=1.4$, $\delta=1$ for
(a) and $\delta=1.4$ for (b). In the calculation, $\omega=1$ has
been taken as units.}
\end{figure}

We note that the G function contains all the information of our
system, including both eigenvalues and eigenfunctions. As the
eigenvalues are determined, coefficients of the corresponding
eigenfunctions can be obtained by recursion relations (\ref{Kn}).
Although the G-function is represented as a summation of infinite
power series, undoubtedly practical calculations of the zeros of G
function require truncations of coherent state orbits.
Nevertheless, the compact form of our solution implies that we can
get desired accurate results in our scheme. A connection with the
quantum Rabi model inspires us to investigate the dynamical
behavior of the Rabi-like model,
which may stimulate the study of
quantum dynamics related to quantum Rabi model by using cold
atomic systems.

In order to study the dynamics of Rabi-like model, we first
prepare the bosonic atoms in a harmonic trap and suppose that the
interaction between atoms is very weak and the system can be
treated as noninteracting bosons. In principle, one can tune the
interaction between atoms to be zero by the Feshbach resonance. We
take the initial state as $|0\rangle_a\otimes\left(1,0\right)^T$,
which is the ground state of the system in a harmonic trap. At
time $T=0$, we turn on the Raman-induced SO coupling. The system
will evolve under the Hamiltonian (\ref{Rabi-like}) and the
time-dependent wavefunction reads
\begin{eqnarray}
\left| {\psi(t)} \right\rangle = e^{-i\hat{H}t} \left|{\psi(0)}
\right\rangle  = \sum_{n} e^{-i E_n t} \left|
{\psi_n}\right\rangle \frac{ \langle \psi_n \left| \psi(0)
\right\rangle} {\langle {\psi_n} \left| {\psi_n} \right\rangle},
\end{eqnarray}
where $\left| {\psi_n} \right\rangle$ is the eigenstate of the
Rabi-like model corresponding to the eigenvalue $E_n$. The
analytical solution of the Rabi-like model enables us to calculate
the time evolution of the expectation value of $\sigma_z$ in a
numerically exact way \cite{dynamics,dynamics2}.

The expectation value of $\sigma_z$ is defined as $P_z(t) = \left
\langle {\psi(t)} \right|  \sigma_z \left| {\psi(t)}
\right\rangle$, which describes the population difference of two
component atoms. The left panels of Fig.3 show the time-dependent
population difference $P_z(t)$ for $\Omega=0.5$, $\delta=0$ under
various coupling strengths. For the weak coupling strength with
$g=0.1$, $P_z(t)$ oscillates between $1$ and $-1$, which indicates
the periodic population swapping between different component
atoms. With increasing the coupling strength, the typical Rabi
oscillation in the weak coupling regime breaks down. For $g=1.5$,
$P_z(t)$ displays a sharp steplike decay with similar dynamic
behavior as in the quantum Rabi model \cite{dynamics}. In the
right panels of Fig.3, we show the time-dependent population
difference for system with $\Omega=0.5$ and $\delta=0.4$. It is
obvious that the oscillation amplitude of $P_z(t)$ is suppressed
by the nonzero term of $\delta$. From the Hamiltonian
(\ref{Rabi-like}), given the initial state as $(1,0)^T$ in
$\sigma_z$ representation, one may understand the competition
between the $\Omega$ term and $\delta$ term. While the nondiagonal
term of $\Omega$ flips the spin, the diagonal term of $\delta$
tends to stabilize the initial state. The effect of coupling $g$
is somewhat like the diagonal term while in a more subtle way.

In order to compare with the well-known quantum Rabi model, we
take $\Omega$ with the same energy order as $\omega$, which is
generally fulfilled for cases of quantum optical systems. However,
the energy scale of cold atomic systems is quite different from
quantum optical systems. In quantum optical systems,
$\Omega\approx\omega$ and the coupling $g$ is the smallest
quantity, while in cold atom systems, the harmonic trap is usually
not deep enough and generally $\Omega \gg \omega$. For example, in
the recent experiment of USTC group \cite{ZhangPRA}, one has
$\omega=(2\pi)50Hz$, $E_{recoil}=(2\pi)2.21kHz$,
$\delta=4E_{recoil}$, $\Omega\approx E_{recoil}$, from which we
can get $\Omega\approx40$ and  $\delta\approx160$ in units of
$\omega$. To see the effect of a large $\Omega$, in Fig.4, we give
an example for the case with $\Omega=10$. Comparing with Fig.3, we
can clearly
 see the oscillation frequency increases with increasing the Raman coupling
strength. The oscillation amplitude is strongly suppressed by the
$\delta$ term.

The mapping of the trapped atomic gas with spin-orbit coupling to
quantum Rabi-like model may stimulate the study of interesting
phenomena related to the basic quantum optical model by using the
cold atomic system. Although the parameters of trapped cold atomic
systems in current experiment are much different from the general
Rabi model, the good tunability of cold atomic systems may make it
be possible to access the parameter regime of traditional quantum
Rabi model by increasing the frequency of trap potential. One
possible way is using the 1D deep optical lattice to produce a
series of two-dimensional (2D) Bose gases. In the direction of 1D
optical lattice, each 2D gas can be viewed to be tightly trapped
by an effective harmonic trap. By this way, the frequency of the
effective harmonic trap may be tuned to the same order of
Raman-coupling strength \cite{WangRQ}. As quantum Rabi model in
the quantum optical system and SO coupled atomic gases can be
viewed as two limit cases in parameter spaces, one may also
explore some novel properties of the Rabi-like model by using cold
atomic systems in the parameter regime which is not accessible in
traditional optical systems.

\begin{figure}[tbp]
\includegraphics[width=3.5in]{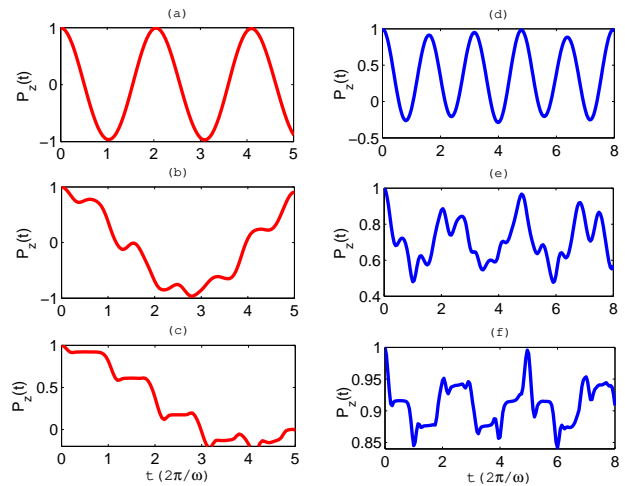}
\caption{(Color online) Expectation value of $\sigma_z$ for
 $\delta=0$ (red line) and $\delta=0.4$ (blue line). From (a) to (c) or (d) to (f), the
coupling strength is taken as $0.1$, $0.7$ and $1.5$,
respectively. Here
 $\Omega=0.5$, time $t$ is in units of $2\pi/\omega$ }
\end{figure}
\begin{figure}[tbp]
\includegraphics[width=3.5in]{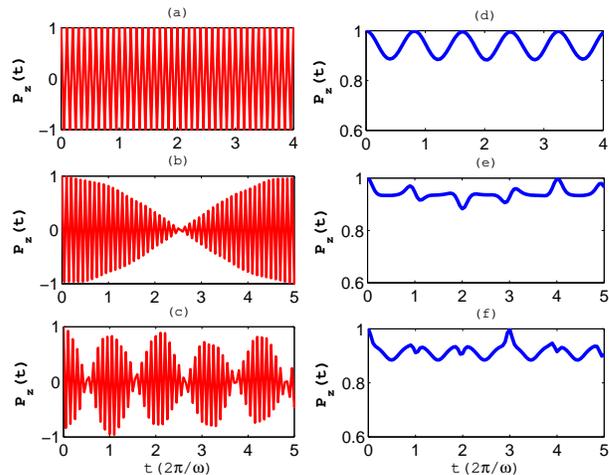}
\caption{(Color online) Expectation value of $\sigma_z$ for
$\delta=0$ (red line) and $\delta=40$ (blue line).
 From (a) to (c) or (d) to (f), the
coupling strength is taken as $0.1$, $0.7$ and $1.5$,
respectively. Here $\Omega=10$, time $t$ is in units of
$2\pi/\omega$ }
\end{figure}

\section{Summary}

In conclusion, we show that the ultracold atomic system in a
harmonic trap with Raman-induced SO coupling is equivalent to a
Rabi-like model and the closed form of the exact solution is
obtained. The connection of the trapped cold atoms with SO
coupling to the Rabi-like model enables us to explore a variety of
properties of the cold atomic system by using traditional methods
in quantum optics. We want to point out that this connection is
rather implicit as the harmonic trap here provide a source of
bosonic field, which was only considered as a confinement in
previous studies. Another reason of this implicity is the
different energy scales in quantum optics and cold atoms. Rapid
progress in cold atomic experiments gives the possibility to
prepare our system in quantum optics limit. Our study also
provides a new vision for understanding the trapped cold atomic
systems with SO coupling.

\begin{acknowledgments}
We thank R. Q. Wang and P. Zhang for helpful discussions. This
work has been supported by National Program for Basic Research of
MOST, NSF of China under Grants No.11121063 and No.11174360 and
973 grant.
\end{acknowledgments}

\appendix
\section{Derivation of solution.} For simplicity, we shall use the
method of extended coherent states developed by Chen et. al.
\cite{chenqh}, which was used to re-derive the Braak's solution to
the quantum Rabi model. Following the procedure for the Rabi model
\cite{chenqh}, we can get the solution of our Rabi-like model.

Left-multiplying $_A\left\langle n\right|$ to the eigenvalue
equation
 $H \left| { \psi }\right\rangle = E \left| { \psi }\right\rangle $,
we can get the following restricted relations to the coefficients
$J_n^-$ and $K_n^-$:
\begin{eqnarray*}
&& \left( n-\alpha -E\right) J_n^-= -\frac \Omega 2 K_n^- ,\\
&& \left( n+\beta -E\right) K_n^- + 2ig \left( n+1\right)
K_{n+1}^- -2ig K_{n-1}^- \\
&&  ~~~~~~~~~~ = -\frac \Omega 2 J_n^- .
\end{eqnarray*}
From the above equations, we can determine the coefficient $f_n$
recursively,
\begin{eqnarray}
J_n^-&=&-\frac{\Omega}{2\left(n-\alpha-E\right)}K_n^-\label{cor1} , \\
n K_n^-&=& f_{n-1}^-  K_{n-1}^- + K_{n-2}^- , \\
f_n^-&=& \frac {-1}{2ig}\left[ \left( n +\beta -E\right)
-\frac{\Omega ^2}{ 4\left( n -\alpha -E\right) }\right]
\end{eqnarray}
with\ $K_0^-=1\;$and $K_1^-= f_0^-$

Similarly, proceeding as before, we get the relations for two
coefficients $K_n^{+}\;$and $J_n^{+}$:
\begin{eqnarray*}
&& \left( n-\alpha ^{\prime} - E \right) J_n^{+} = -\frac \Omega 2K_n^{+}\\
&& \left( n+\beta ^{\prime }-E\right) K_n^{+} + 2ig \left( n+1\right) K_{n+1}^{+} - 2ig K_{n-1}^+\\
&& ~~~~~~~~~~ = -\frac \Omega 2 J_n^{+},
\end{eqnarray*}
then we have
\begin{eqnarray}
J_n^{+}&=&-\frac{\Omega }{2(n-\alpha ^{\prime
}-E)}K_n^{+} \label{cor2}\\
n K_n^{+} &=&f^{+}_{n-1} K_{n-1}^{+}+K_{n-2}^{+},
\\
f^{+}_{n}&=& \frac {-1} {2ig}\left[ \left( n+\beta ^{\prime
}-E\right) -\frac{\Omega ^2}{4\left( n -\alpha ^{\prime }-E\right)
}\right]
\end{eqnarray}
with $K_0^{+}=1\;$ and $\;K_1^{+}=f^{+}_0.$

As alternative representations of a non-degenerate state with the
eigenvalue $E$, the wavefunction (\ref{wave1}) and (\ref{wave2})
should be equivalent, i.e., $ \left| {\psi} \right\rangle= r
\left| {\psi'}\right\rangle$, where $r$ is a constant.
The above equivalent requirement leads to the following relations:
\begin{eqnarray}
\sum_{n=0}^\infty \sqrt{n!}J_n^-\left| n\right\rangle _A
&=&r\sum_{n=0}^\infty
(-1)^n\sqrt{n!}K_n^{+}\left| n\right\rangle _B, \\
\sum_{n=0}^\infty \sqrt{n!}K_n^-\left| n\right\rangle _A
&=&r\sum_{n=0}^\infty (-1)^n\sqrt{n!}J_n^{+}\left|
n\right\rangle _B.
\end{eqnarray}
Left multiplying the vacuum state $_{a}{\langle }0|$ to both sides
of the above equations and making use of the relation of $
\sqrt{n!}_{a}{\langle }0|n{\rangle }_A=(-1)^n\sqrt{n!}_{a}{\langle }0|n{\rangle }%
_B=e^{-g^2/2}(-ig)^n$, we can eliminate the ratio constant $r$ and
get the following relation
\begin{eqnarray}
& & \sum_{n=0}^\infty J_n^-(-ig)^n\sum_{n=0}^\infty J_n^{+}(-ig)^n \nonumber \\
&=& \sum_{n=0}^\infty K_n^-(-ig)^n\sum_{n=0}^\infty
K_n^{+}(-ig)^n.
\end{eqnarray}
Making substitutions in terms of Eqs. (\ref{cor1}) and
(\ref{cor2}), we obtain
\begin{eqnarray}
&&\sum_{n=0}^\infty \frac{\Omega /2}{n-\alpha
-E}K_n^-(-i g)^n\sum_{n=0}^\infty
\frac{\Omega /2}{n-\alpha ^{\prime }-E}K_n^{+}(-i g)^n  \nonumber \\
&=&\sum_{n=0}^\infty K_n^-(-i g)^n\sum_{n=0}^\infty K_n^{+}(-i g)^n.
\end{eqnarray}
In terms of the the transcendental function $G_\delta (x)$ defined
by Eq.(\ref{bias}) in the main text, the above equation is nothing
else but $G_\delta (x) =0$ with $x=E+g^2$.


\begin{thebibliography}{99}
\bibitem{NIST} Y.-J. Lin, R. L. Compton, A. R. Perry, W. D. Phillips, J. V.
Porto, and I. B. Spielman, Phys. Rev. Lett. \textbf{102}, 130401 (2009);
Y.-J. Lin, R. L. Compton, K. Jim\'{e}nez-Garc\'{\i}a, J. V. Porto, and I. B.
Spielman, Nature \textbf{462}, 628 (2009).

\bibitem{NIST_elec} Y.-J. Lin, R. L. Compton, K. Jim\'{e}nez-Garc\'{\i}a, W.
D. Phillips, J. V. Porto, and I. B. Spielman, Nature Physics \textbf{7}, 531
(2011).

\bibitem{SOC} Y.-J. Lin, K. Jim\'{e}nez-Garc\'{\i}a, and I. B. Spielman,
Nature \textbf{471}, 83 (2011).


\bibitem{JingPRA} Z. Fu, P. Wang, S. Chai, L. Huang, and J. Zhang, Phys.
Rev. A {\bf 84}, 043609 (2011).

\bibitem{collective_SOC} J.-Y. Zhang, S.-C. Ji, Z. Chen, L. Zhang, Z.-D. Du, B. Yan, G.-S. Pan, B. Zhao, Y. Deng, H. Zhai, S. Chen,
and J.-W. Pan, Phys. Rev. Lett. {\bf 109}, 115301 (2012).

\bibitem{NIST_partial} R. A. Williams, L. J. LeBlanc, K. Jim\'{e}nez-Garc%
\'{\i}a, M. C. Beeler, A. R. Perry, W. D. Phillips, and I. B. Spielman,
Science \textbf{335}, 314 (2012).

\bibitem{ZhangPRA} L. Zhang, et. al.,  Phys. Rev. A {\bf 87}, 011601(R)
(2013).

\bibitem{SOC_Fermi} P. Wang, Z.-Q. Yu, Z. Fu, J. Miao, L. Huang, S. Chai, H.
Zhai, and J. Zhang,  Phys. Rev. Lett. {\bf 109}, 095301 (2012).

\bibitem{SOC_MIT} L. W. Cheuk, A. T. Sommer, Z. Hadzibabic, T. Yefsah, W. S.
Bakr, and M. W. Zwierlein,  Phys. Rev. Lett. {\bf 109}, 095302
(2012).



\bibitem{gauge_review} J. Dalibard, F. Gerbier, G. Juzeli\={u}nas, and P.
\"{O}hberg, Rev. Mod. Phys. \textbf{83}, 1523 (2011).

\bibitem{You} Z. F. Xu and L. You, Phys. Rev. A {\bf 85}, 043605
(2012).


\bibitem{Victor} T. D. Stanescu, B. Anderson, and V. Galitski, Phys. Rev. A
\textbf{78}, 023616 (2008).

\bibitem{Stripe} C. Wang, C. Gao, C.-M. Jian, and H. Zhai, Phys. Rev. Lett.
\textbf{105}, 160403 (2010).

\bibitem{Ho} T.-L. Ho, and S. Zhang, Phys. Rev. Lett. \textbf{107}, 150403
(2011).

\bibitem{Cui} X. Cui and Q. Zhou, arXiv:1206.5918; Q. Zhou and X.
Cui,  arXiv:1210.5853.

\bibitem{Wu} C.-J. Wu, I. Mondragon-Shem, and X.-F. Zhou, Chin. Phys. Lett.
\textbf{28} 097102 (2011).

\bibitem{Hu} H. Hu, B. Ramachandhran, H. Pu, and X.-J. Liu, Phys. Rev. Lett.
\textbf{108}, 010402 (2012).

\bibitem{Santos} S. Sinha, R. Nath, and L. Santos, Phys. Rev. Lett. \textbf{%
107}, 270401 (2011).

\bibitem{Ghosh} S. K. Ghosh, J. P. Vyasanakere, and V. B. Shenoy, Phys. Rev. A
\textbf{84}, 053629 (2011).

\bibitem{Larson} J. Larson, B. M. Anderson, and A. Altland,  Phys. Rev. A
\textbf{87}, 013624 (2013).

\bibitem{Rabi} I. I. Rabi, Phys. Rev. \textbf{49}, 324 (1936); \textbf{51}, 652 (1937).


\bibitem{JC}  E. T. Jaynes and F. W. Cummings, Proc. IEEE, \textbf{
51}, 89 (1963).


\bibitem{Braak}  D. Braak,  Phys. Rev. Lett. \textbf{107}, 100401
(2011).

\bibitem{chenqh}  Q. H. Chen, C. Wang, S. He, T. Liu and K. L. Wang, Phys. Rev. A \textbf{86},
023822 (2012).

\bibitem{Ziegler} K. Ziegler, J. Phys. A, \textbf{45}, 452001
(2011).

\bibitem{Solano}  E. Solano, Physics 4, 68 (2011).

\bibitem{Gardas} B. Gardas and J. Dajka, arXiv:1301.5660.

\bibitem{dynamics} F. A. Wolf, M. Kollar, and D.
Braak, Phys. Rev. A \textbf{85}, 053817 (2012).



\bibitem{FengM} T. Liu, K. L. Wang, and M. Feng, Europhys. Lett. \textbf {86}, 54003
(2009).

\bibitem{chenqh-EPL}  Q. H. Chen, T.  Liu, Y. Y. Zhang,  and K. L. Wang,  Europhys. Lett. \textbf{96}, 14003
(2011).

\bibitem{CQED}  J. M. Raimond, M. Brune, and S. Haroche, Rev. Mod. Phys.
\textbf{73}, 565 (2001);  H. Mabuchi and A. C. Doherty, Science
\textbf{298}, 1372 (2002).


\bibitem{CQED2} T. Niemczyk, et al., Nature Physics \textbf{6},
772 (2010); P. Forn-D\'{i}az et al., Phys. Rev. Lett.
\textbf{105}, 237001 (2010).

\bibitem{Semiconducor} K. Hennessy, et al., Nature (London) {\textbf 445},
896 (2007).

\bibitem{dynamics2} Y.-Y. Zhang, Q.-H. Chen, and S.-Y. Zhu,
arXiv:1106.2191.

\bibitem{WangRQ} R. Q. Wang, private communication.

\end{thebibliography}
\end{document}